\documentclass[a4paper,12pt]{article}
\usepackage{latexsym}
\usepackage{rotating}
\usepackage{amsmath}
\usepackage{amsfonts}
\usepackage{amssymb}
\author{
\small Wen-Xiu Ma\\
\small 
Department of Mathematics, Zhejiang Normal University, \\
\small Jinhua
321004, People’s Republic of China
\\
\small College of Mathematics and Systems Science, Shandong University of Science and Technology, \\
\small Qingdao 266590, Shandong,  China
\\
\small Department of Mathematics and Statistics, University of South Florida, \\
\small
Tampa, FL 33620, USA\\
\small Department of Mathematical Sciences, North-West University, 
\\
\small Mafikeng Campus, Private Bag X 2046, Mmabatho 2735, South Africa
}
\title {Conservation laws 
by symmetries and adjoint symmetries}
\date{\nonumber}
\begin{document}

\setlength{\baselineskip}{15.6pt}
\maketitle

\begin{abstract}

Conservation laws are formulated for systems of differential equations by using 
symmetries and adjoint symmetries, 
and an application to systems of evolution equations
is made, together with illustrative examples.
The formulation does not require the existence of a Lagrangian for a given system, and 
the presented examples include computations of conserved densities for the heat equation,
Burgers' equation and the Korteweg-de Vries equation.

\end{abstract}

\newcommand{\Z}{\mathbb{Z}}
\newcommand{\C}{\mathbb{C}}
\newcommand{\R}{\mathbb{R}}

\def \ba {\begin{array}}
\def \ea {\end{array}}
\def\bea{\begin{eqnarray}}
\def\eea{\end{eqnarray}}
\def \part {\partial }

\def \D {\displaystyle{}}

\renewcommand{\theequation}{\thesection.\arabic{equation}}

\newtheorem{lemma}{Lemma}[section]
\newtheorem{theorem}{Theorem}[section]
\newtheorem{definition}{Definition}[section]

%need to change
\def \be  {\begin{equation}}
\def \ee  {\end{equation}}
\def \al   {\alpha }
\def \la    {\lambda  }

\section{Introduction}
\setcounter{equation}{0}

For systems of Euler-Lagrange equations,
Noether's theorem shows that symmetries lead to 
conservation laws
\cite{Bluman-book1989,Olver-book1986}. 
The Lagrangian formulation of the systems under consideration 
is essential in presenting conservation laws from symmetries, and
many physically important examples can be found in   
\cite{Bluman-book1989}-\cite{Morawetz-BAMS2000}.
Is it possible to extend such a connection 
between conservation laws and symmetries for systems of non-Euler-Lagrange equations?
A definite answer has been given in the case of systems of discrete evolution equations \cite{Ma-Symmetry2015}.
We would, in this paper, like to
formulate a similar theory of conservation laws 
for systems of differential equations and apply it to systems of evolution equations. 
More precisely, we want to exhibit that 
pairs of symmetries and adjoint symmetries lead to conservation laws
for whatever systems of differential equations.

For totally nondegenerate systems of differential equations,
conservation laws are classified by characteristic forms \cite{Olver-book1986}.
The characteristics, also called the multipliers \cite{AncoB-PRL1997}, of conservation laws are adjoint symmetries, and thus, the existence of an adjoint symmetry is necessary for a 
totally nondegenerate system of differential equations to admit a 
 conservation law. 
Adjoint symmetries generate conservation laws, when the variational derivatives of product functionals of the adjoint symmetries and the systems under consideration vanish \cite{AncoB-PRL1997}. For systems of evolution equations, 
functionals 
are 
conserved if and only if 
their variational derivatives are adjoint symmetries \cite{Ma-Symmetry2015,MaZ-NJMP2002}. 
Nonlinear self-adjointness has also been introduced on the basis of adjoint systems to construct conservation laws for nonlinearly self-adjoint systems of differential equations, where the nonlinear self-adjointness means that the second set of dependent variables in an adjoint system stands for an adjoint symmetry \cite{Ibragimov-JMAA2007,Ibragimov-JPA2011}.
We will directly utilize both symmetries and adjoint symmetries to produce conservation laws for systems of differential equations, without using 
any Lagrangian or extended Lagrangian. 

This paper is organized as follows. In section 2, a formulation of conservation laws is furnished
for systems of differential equations, regardless of the existence of a Lagrangian.
In section 3, an application is made for systems of evolution equations, and in section 4, the resulting theory is 
 used to compute conserved densities for
the heat equation, Burgers' equation and the Korteweg-de Vries equation, along with many new 
conserved densities. Finally in section 5, a few of concluding remarks are 
given with some discussion.
 
\section{A formulation of conservation laws}
\setcounter{equation}{0}

We will use a plain language to formulate the correspondence between conservation laws and pairs of symmetries and adjoint symmetries, 
although there is a geometrical language.

Let $x=(x^1,\cdots,x^p)\in \R ^p$, $u=(u^1,\cdots,u^q)^T, 
\ u^i=u^i(x),\ 1\le i\le q$, and
\be u^i_\al =D^\al u^i,\ D^\al = \part _1^{\al _1}\cdots \part _p^{\al _p},\ 
\part _j=\frac {\part }{\part x^j},\  1 \le i\le q,\ 1\le j\le p,\ee 
for $\al =(\al _1,\cdots ,\al _p)$ 
with non-negative integers $\al _i$, $1\le i\le p$. 
Assume that   
 ${\cal A}$
 denotes the space of all functions $f(x,u,\cdots,u^{(n)})$ with $n\ge 0$,
where $f$ is a smooth function of the involved variables 
and $u^{(n)}$ is the set of $n$-th order partial derivatives of $u$
with respect to $x$,
and 
${\cal B}$
denotes the space of all smooth functions 
of $x,u$ and derivatives of $u$ with respect to $x$
to some finite order.
The locality
 means that a function $f(u)$ 
depends locally on 
$u$ with respect to $x$, i.e., 
any value $(f(u))(x)$ is completely determined by the value of $u$ 
in a sufficiently small region of $x$.
 Any functions 
in ${\cal A}$, particularly
differential polynomial functions, are local.
The space ${\cal B}$ contains nonlocal functions,
and simple examples are 
$x^l\int _0^{x^k}\part _j u^i\,dx^k,\ j\ne k $.
We use ${\cal A}^r$ and
${\cal B}^r$ to denote the $r$-th order tensor products of 
${\cal A}$ and ${\cal B}$:
\be
{\cal A}^r
= \{ ( f_1,\cdots, f_r)^T
  | \, f_i\in {\cal A},\ 1\le i\le r  \}
,\ 
{\cal B}^r=
 \{( f_1,\cdots, f_r)^T
  | \, f_i\in {\cal B},\ 1\le i\le r  \}
.
\ee

Two functions $f_1,f_2\in {\cal B}$ are said to be equivalent and denoted by $f_1\sim f_2$, if 
there exist $g_i\in {\cal B},\ 1\le i\le p$, such that 
\[ f_1-f_2=\sum_{i=1}^p\partial_i g_i.\]
This is an equavalence relation. Each equivalence class is called a functional and the class that $P\in {\cal B}$ belongs to is denoted by $\int P  \,dx$.
The inner products are defined by  
\[ \langle X, Y\rangle =\int \sum_{i=1}^s X_iY_i\,dx
,\ X=(X_1,\cdots,X_s)^T,\ Y=(Y_1,\cdots,Y_s)^T\in {\cal B}^s,\ s\ge 1
,\] 
and the adjoint operator $\Phi^*: {\cal B}^s\to {\cal B}^r$ of a linear operator $\Phi: {\cal B}^r \to {\cal B}^s$ is determined by
\be 
\langle X,\Phi^* Y\rangle =\langle \Phi X, Y \rangle ,\ X\in {\cal B}^r,\ Y\in {\cal B}^s.
\label{eq:innerproducts:ma-conservationlaws} 
\ee 

For any vector function
$X=X(u)=(X_1,\cdots ,X_r)^T\in {\cal A}^r$, we introduce its Gateaux operator 
\be X'=X'(u)=(V_j(X_i))_{r\times q}=
\left( \ba {cccc} V_1(X_1)&V_2(X_1)&\cdots &V_q(X_1)\vspace{2mm}\\
V_1(X_2)&V_2(X_2)&\cdots &V_q(X_2)\vspace{2mm}\\
\vdots &\vdots &\ddots &\vdots \vspace{2mm}\\
V_1(X_r)&V_2(X_r)&\cdots &V_q(X_r)\ea 
\right),\ V_i(X_j)=\sum _{\al \ge 0}
\frac {\part X_j}{\part u^i_\al }D^\al ,
  \ee 
where $\al \ge 0$ means that all components $\al _i\ge 0,\ 1\le i\le p$.
The adjoint operator $(X^{'})^*$ of $X'$ is given by  
\be (X^{'})^*=(X^{'})^*(u)
=(V_i^* (X_j))_{q\times r},\ V_i^*(X_j) =
\sum _{\al \ge 0}(-D)^\al \frac {\part X_j }{\part u^i_\al }, 
\label{eq:adjointoperatorforX:ma-conservationlaws}
 \ee  
where $(-D)^\al =(-\part _1)^{\al _1}
\cdots (-\part _p)^{\al _p}.$

Now, for a given integer $l\ge 1$, let us consider a system of differential equations 
\be \Delta(u)=0,\ \Delta =(\Delta_1,\cdots,\Delta_l)^T\in {\cal A}^l. 
\label{eq:gsystemofDEs:ma-conservationlaws}\ee 
Its linearized system and adjoint linearized system are defined by
\bea && \Delta'(u)\sigma(u) =0,\ 
\sigma \in {\cal B}^q,\label{eq:glinearizedeqn:ma-conservationlaws} \\
&& (\Delta^{'})^*(u)\rho(u)=0, \ \rho \in {\cal B}^l,\label{eq:galinearizedeqn:ma-conservationlaws}
\eea
respectively. Here $\Delta'$ and $(\Delta^{'})^*$ denote the Gateaux operator of $\Delta$ and its adjoint operator, respectively. It is easy to observe that 
\[ \Delta '(u)\sigma (u) =\Delta '(u)[\sigma (u)]:=\frac {\partial }{\partial \varepsilon }
\Bigl.\Bigr|_{\varepsilon =0} (\Delta (u+\varepsilon \sigma (u))).\]

\begin{definition}
A vector function $\sigma \in {\cal B}^q$ 
is called a symmetry of 
(\ref{eq:gsystemofDEs:ma-conservationlaws}),
if it satisfies  
(\ref{eq:glinearizedeqn:ma-conservationlaws}) when $u$ solves (\ref{eq:gsystemofDEs:ma-conservationlaws}).
A vector function $\rho \in {\cal B}^l $ is called an adjoint symmetry of 
(\ref{eq:gsystemofDEs:ma-conservationlaws}),
if it satisfies 
(\ref{eq:galinearizedeqn:ma-conservationlaws})
 when $u$ solves (\ref{eq:gsystemofDEs:ma-conservationlaws}).
\end{definition}

\begin{definition}
If a total divergence relation 
\be \textrm{Div} \,P=\sum_{i=1}^p\part _iP_i=0,\ P=(P_1,\cdots,P_p)^T\in {\cal B}^p, 
\label{eq:cl:ma-conservationlaws}\ee
holds for all solutions of (\ref{eq:gsystemofDEs:ma-conservationlaws}),
then (\ref{eq:cl:ma-conservationlaws}) is called a conservation law and $P$ a conserved vector of 
(\ref{eq:gsystemofDEs:ma-conservationlaws}).
\end{definition}

Conservation laws include familiar concepts of conservation of mass, energy and momentum arising in physical applications.
In the case of Laplace's equation, 
the equation itself is a conservation law, since
\[\Delta u =  \textrm{Div} \, (
\nabla u)=0\]
 for all solutions.
For a system of ordinary differential equations involving a single independent variable $x\in \R$, a conservation law 
$D_xP=0$ requires that $P(x,u^{(u)})$ be constant for all solutions of the system. Thus, a conservation law for a system of ordinary differential equations is equivalent to the classical notion of a first integral or constant of motion of the system. 

In a dynamical problem, one of the independent variables is distinguished as the time $t$, the remaining variables $x=(x^1,\cdots,x^p)$ being spatial variables. Then, a conservation law 
takes the form 
\[  D_t T +\textrm{Div}\, X =0, \] 
in which $\textrm{Div}$ stands for the spatial divergence of $X\in {\cal B}^p$ with respect to $x$.
Here $T$ is called a conserved density and $X$, the associated flux vector corresponding to $T$.

Conservation laws of a totally nondegenerate system of differential equations 
(\ref{eq:gsystemofDEs:ma-conservationlaws})
are classified through characteristic forms \cite{Olver-book1986}:
\be \textrm{Div} \,P=Q^T \Delta,  \ee 
where $Q\in {\cal B}^l$. 
Such a vector function $Q$ is
 called the characteristic of 
the associated conservation law.  
For a normal, totally nondegenerate system of differential equations, 
equivalent conservation laws correspond to equivalent characteristics \cite{Olver-book1986}. 

We would like to formulate conservation laws 
for 
a general system of differential equations
(\ref{eq:gsystemofDEs:ma-conservationlaws}) by using pairs of  symmetries and 
adjoint symmetries. 
In our expressions of construction,
we will use the following assumption for brevity:
an empty product of derivatives $\part _i^{\al _i}$, 
$\al _i \ge 0,$ $1\le i\le p$, 
is understood to be the identity operator. For example,
$\Pi_{i=r}^s\part _i^2$ 
implies the identity operator, when $r>s$.

\begin{lemma}\label{le:basiceq:ma-conservationlaws}
Let $f$ and $g$ be two smooth functions in variables $x^1,\cdots,x^p$.
Then for any $\al =(\al _1,\cdots,\al _p)$ with non-negative integers
$\al _i$, $1\le i\le p$, we have 
\bea &&
f(D^\al g)-((-D)^\al f)g=f(\partial _1^{\al _1} \cdots \partial _p^{\al _p}g)-
( (-\partial _1)^{\al _1} \cdots (-\partial _p)^{\al _p}f)g  \nonumber \\
&=&
\sum_{i=1}^p\partial _i\sum_{\beta _i=0}^{\al _i-1}
((-\partial _1)^{\al _1} \cdots (-\partial _{i-1})^{\al _{i-1}}(-\partial _i)^{\beta _i}f)
(\partial _i^{\al _i-\beta _i-1}\partial _{i+1}^{\al _{i+1}} \cdots \partial _p^{\al _p}g)
, \qquad 
\label{eq:basiceq:ma-conservationlaws}
\eea
where an empty sum is understood to be zero. 
\end{lemma}

{\it Proof:} 
First note that we have 
\be 
a \partial _i^{k}b-((-\partial _i)^{k}a)b=
\partial _i\sum_{l=0}^{k-1} ((-\partial _i)^{l}a)
(\partial _i^{k-l-1}b),
\  k\ge 1,\ 1\le i\le p,\label{eq:basicbasiceq:ma-conservationlaws}
\ee 
for any two smooth functions $a$ and $b$ in variables $x^1,\cdots,x^p$,
and then we decompose that  
\bea &&
f(D^\al g)-((-D)^\al f)g=
f(\partial _1^{\al _1} \cdots \partial _p^{\al _p}g)-
( (-\partial _1)^{\al _1} \cdots (-\partial _p)^{\al _p}f)g  \nonumber \\
&=&
\sum_{i=1}^p[ ((-\partial _1)^{\al _1} \cdots (-\partial _{i-1})^{\al _{i-1}}f)
(\partial _i^{\al _i} \cdots \partial _p^{\al _p}g)-
 ((-\partial _1)^{\al _1} \cdots (-\partial _i)^{\al _i}f)
(\partial _{i+1}^{\al _{i+1}} \cdots \partial _p^{\al _p}g)].\nonumber \eea
It follows from (\ref{eq:basicbasiceq:ma-conservationlaws}) that each term in the above sum
can be computed as follows:
\bea && 
((-\partial _1)^{\al _1} \cdots (-\partial _{i-1})^{\al _{i-1}}f)
(\partial _i^{\al _i} \cdots \partial _p^{\al _p}g
-
 ((-\partial _1)^{\al _1} \cdots (-\partial _i)^{\al _i}f)
(\partial _{i+1}^{\al _{i+1}} \cdots \partial _p^{\al _p}g) 
\nonumber \\
&=&((-\partial _1)^{\al _1} \cdots (-\partial _{i-1})^{\al _{i-1}}f)
\partial _i^{\al _i}(\partial _{i+1}^{\al _{i+1}}  \cdots \partial _p^{\al _p}g)
\nonumber \\ && 
- ((-\partial _i)^{\al _i}((-\partial _1)^{\al _1} \cdots (-\partial _{i-1})^{\al _{i-1}}
f)) (\partial _{i+1}^{\al _{i+1}} \cdots \partial _p^{\al _p}g) 
\nonumber \\
&=&
\part _i\sum_{\beta_i=0}^{\al _i-1} 
((-\partial _1)^{\al _1} \cdots (-\partial _{i-1})^{\al _{i-1}}(-\partial _i)^{\beta _i}f)
(\partial _i^{\al _i-\beta _i-1}\partial _{i+1}^{\al _{i+1}} \cdots \partial _p^{\al _p}g),
\ 1\le i\le p,
\nonumber \eea
where an empty sum is understood to be zero. 
This allows us to conclude that our equality
(\ref{eq:basiceq:ma-conservationlaws}) holds
for any $\al =(\al _1,\cdots,\al _p)$.
The proof is finished.
$\vrule width 1mm height 3mm depth 0mm$

This lemma tells us that $fD^\al g-((-D)^\al f)g$ is a total divergence function
for any two smooth functions $f$ and $g$, and also guarantees that \eqref{eq:adjointoperatorforX:ma-conservationlaws} presents an adjoint operator of $X'$.

\begin{theorem}\label{thm:gclstructure:ma-conservationlaws}
Let $\sigma =(\sigma _1,\cdots,\sigma_q)^T\in {\cal B}^q$ and $\rho 
=(\rho _1,\cdots,\rho_l)^T\in {\cal B}^l$ 
be a symmetry and an adjoint symmetry of 
a system of differential equations (\ref{eq:gsystemofDEs:ma-conservationlaws}), respectively.
Then we have a conservation law for the system
(\ref{eq:gsystemofDEs:ma-conservationlaws}):
\be
 \sum_{k=1}^p\partial _k
\sum_{i=1}^l \sum_{j=1}^q \sum_{\al  \ge 0}
\sum_{\beta _k=0}^{\al _k-1}
((-\partial _1)^{\al _1} \cdots (-\partial _{k-1})^{\al _{k-1}}(-\partial _k)^{\beta _k}
\rho_i\frac{\part \Delta_i}{\part u^j_\al })
(\partial _k^{\al _k-\beta _k-1}\partial _{k+1}^{\al _{k+1}} \cdots \partial _p^{\al _p}
\sigma _j)=0, 
\label{eq:gcl:ma-conservationlaws}
\ee 
where $\al =(\al _1,\cdots,\al _p)$, 
and an empty sum is understood to be zero. 
\end{theorem}

{\it Proof:}
Let us compute that 
\bea  
&&
  \rho^T \Delta '\sigma - \sigma ^T(\Delta ^{'})^* \rho  
= \sum_{i=1}^l  \sum_{j=1}^q ( \rho_iV_j(\Delta _i)\sigma_j- \sigma _j V_j^* (\Delta _i)
\rho_i )\nonumber \\ 
& &=
\sum_{i=1}^l\sum_{j=1}^q \sum_{\al \ge 0} ( \rho_i\frac {\part \Delta_i}{\part u^j_\al }D^\al \sigma _j - 
\sigma _j (-D)^\al \rho_i \frac {\part \Delta _i}{\part u^j_\al } ).
\nonumber \eea
By using Lemma \ref{le:basiceq:ma-conservationlaws},
for all $ 1\le i\le l, 1\le j\le q$ and $ \al \ge 0,$ we have 
\bea &&
 \rho_i\frac {\part \Delta _i}{\part u^j_\al }D^\al \sigma _j - 
\sigma _j (-D)^\al \rho_i \frac {\part \Delta _i}{\part u^j_\al } 
=
\rho_i\frac {\part 
\Delta _i}{\part u^j_\al }(D^\al \sigma _j )- 
((-D)^\al \rho_i \frac {\part \Delta _i}{\part u^j_\al })\sigma_j
\nonumber \\
&=& 
 \sum_{k=1}^p\partial _k\sum_{\beta _k=0}^{\al _k-1}
((-\partial _1)^{\al _1} \cdots (-\partial _{k-1})^{\al _{k-1}}(-\partial _k)^{\beta _k}
\rho_i\frac{\part \Delta _i}{\part u^j_\al })
(\partial _k^{\al _k-\beta _k-1}\partial _{k+1}^{\al _{k+1}} \cdots \partial _p^{\al _p}
\sigma _j), \nonumber
\eea
where an empty sum is understood to be zero.
Now, 
noting that   
$\Delta '\sigma =0$ and $(\Delta ')^*\rho =0$ hold for all solutions of (\ref{eq:gsystemofDEs:ma-conservationlaws}),
we see that (\ref{eq:gcl:ma-conservationlaws}) follows. 
The proof is finished.
$\vrule width 1mm height 3mm depth 0mm$

The theorem gives us an explicit formulation of conservation laws 
for systems of differential equations, regardless of the existence of a Lagrangian. For totally nondegenerate systems of differential equations, we can have
\be 
\Delta ' \sigma = R_\sigma^{\textrm{sym}} \Delta ,\ (\Delta ')^*\rho  =R_\rho^{\textrm{asym}}  \Delta ,
\ee
where $R_\sigma^{\textrm{sym}}  $  and 
$R_\rho^{\textrm{asym}} $ 
are two $l\times l$ and $q\times l$ matrix differential operators depending on 
$\sigma $ and $\rho$, respectively \cite{Olver-book1986}, 
and then 
the following computation 
\be 
\textrm{Div}\, P=
\rho^T \Delta '\sigma - \sigma ^T(\Delta ^{'})^* \rho  
=\rho^T  R_\sigma^{\textrm{sym}} \Delta -\sigma ^T
R_\rho^{\textrm{asym}}  \Delta  
\sim (
(R_\sigma^{\textrm{sym}})^*\rho - (R_\rho^{\textrm{asym}} )^*\sigma
) ^T\Delta 
\ee
presents the characteristic of an equivalent conservation law:
\be 
Q=(R_\sigma^{\textrm{sym}})^*\rho - (R_\rho^{\textrm{asym}} )^*\sigma. 
\label{eq:Qforsigmaandrho:ma-conservationlaws}
\ee

A direct way to generate more conservation laws from known ones 
is to use 
recursion structures. 
Similar to the definition of recursion 
operators \cite{Olver-JMP1978},
 or hereditary symmetry operators 
\cite{Fuchssteiner-NA1979}, 
we can also have recursion structures of other kinds for systems of differential equations.
Recursion operators transforms symmetries to symmetries, and hereditary symmetry operators provide recursion operators for hierarchies. 

\begin{definition}
If an operator $\Psi (u)$ transforms an adjoint symmetry  of 
(\ref{eq:gsystemofDEs:ma-conservationlaws}) 
into another adjoint symmetry of (\ref{eq:gsystemofDEs:ma-conservationlaws}),
then $\Psi (u)$ is called an adjoint recursion operator
of 
(\ref{eq:gsystemofDEs:ma-conservationlaws}).
\end{definition}

\begin{definition}
If an operator $\bar \Phi (u)$ transforms an adjoint symmetry 
of 
(\ref{eq:gsystemofDEs:ma-conservationlaws}) 
into a symmetry of 
(\ref{eq:gsystemofDEs:ma-conservationlaws}) ,
then $\bar \Phi (u)$ is called a Noether operator
of 
(\ref{eq:gsystemofDEs:ma-conservationlaws}).
Conversely, if an operator $\bar \Psi (u)$ transforms a symmetry 
of 
(\ref{eq:gsystemofDEs:ma-conservationlaws}) 
into an adjoint symmetry of (\ref{eq:gsystemofDEs:ma-conservationlaws}),
then $\bar \Psi (u)$ is called an inverse Noether operator
of 
(\ref{eq:gsystemofDEs:ma-conservationlaws}).
\end{definition} 

All the above operators are important in establishing 
more conservation laws, and thus
integrability of 
systems of differential equations.  
The concepts of Noether operators
and inverse Noether operators were also introduced for integrable systems
 \cite{FuchssteinerF-PD1981}. 

\section{Application to evolution equations}
\setcounter{equation}{0}

\subsection{Conservation laws}

Let us take
a set of independent variables $(t,x^1,\cdots,x^p)$,
including a 
distinguished time variable $t\in \R$, 
 and consider
a system of evolution equations 
\be u_t=K(u),\  K\in {\cal A}^q. \label{eq:systemofEEs:ma-conservationlaws}\ee 
Obvisouly, we have 
$\Delta =u_t-K(u)$ with $l=q$, and thus
its linearized system and adjoint linearized system read 
\bea &&(\sigma(u)) _t=K'(u)\sigma(u) ,\ 
\sigma \in {\cal B}^q,\label{eq:linearizedeqn:ma-conservationlaws} \\
&& (\rho(u)) _t=-(K^{'})^*(u)\rho(u), \ \rho \in {\cal B}^q,\label{eq:alinearizedeqn:ma-conservationlaws}
\eea
respectively. Here $K'$ and $(K^{'})^*$ stand for the Gateaux operator of $K$ and its adjoint 
operator, respectively.

It is easy to see that 
two vector functions $\sigma ,\rho\in {\cal A}^q$ are 
 a symmetry and an adjoint symmetry of the system 
(\ref{eq:systemofEEs:ma-conservationlaws})
if and only if they satisfy 
\bea && 
\frac {\part \sigma (u)}{\part t}=K'(u)\sigma (u)- \sigma' (u)K(u),
\label{eq:equivalentconditionofsymmetries}
\\ &&
\frac {\part \rho (u)}{\part t}=-(K^{'})^*(u)\rho (u)- \rho' (u)K(u),
\label{eq:equivalentconditionofadjointsymmetries} 
\eea
respectively,
when $u$ solves (\ref{eq:systemofEEs:ma-conservationlaws}),
where $\sigma'$ and $\rho'$ are the Gateaux operators of $\sigma $ and $\rho $.

A total divergence 
\be T_t=\sum_{i=1}^p\part _i X_i,\ T,X_i\in {\cal B},\ 1\le i\le p, 
\label{eq:clofees:ma-conservationlaws}\ee
gives us a conservation law
for the system of evolution equations (\ref{eq:systemofEEs:ma-conservationlaws}),
and $T$ is a
conserved density of (\ref{eq:systemofEEs:ma-conservationlaws})
and $X=(X_1,\cdots,X_p)^T$, a conversed flux vector of (\ref{eq:systemofEEs:ma-conservationlaws})
corresponding to $T$.

An application of 
Theorem \ref{thm:gclstructure:ma-conservationlaws} 
to systems of evolution equations presents 
the following result on conservation laws for systems of evolution equations.

\begin{theorem}\label{thm:clstructureforees:ma-conservationlaws}
Let $\sigma =(\sigma _1,\cdots,\sigma_q)^T\in {\cal B}^q$ and $\rho 
=(\rho _1,\cdots,\rho_q)^T\in {\cal B}^q$ 
be a symmetry and an adjoint symmetry of 
a system of evolution equations (\ref{eq:systemofEEs:ma-conservationlaws}), respectively.
Then we have a dynamical conservation law for the system  
(\ref{eq:systemofEEs:ma-conservationlaws}):
\be
(\sigma ^T\rho )_t=(\sum_{i=1}^q\sigma_i\rho_i )_t =
\sum_{k=1}^p \partial _k X_k ,
\label{eq:clofees:ma-conservationlaws}
\ee
where the conserved fluxes are defined by 
\be 
X_k=
\sum_{i,j=1}^q \sum_{\al  \ge 0}
\sum_{\beta _k=0}^{\al _k-1}
((-\partial _1)^{\al _1} \cdots (-\partial _{k-1})^{\al _{k-1}}(-\partial _k)^{\beta _k}
\rho_i\frac{\part K_i}{\part u^j_\al })
(\partial _k^{\al _k-\beta _k-1}\partial _{k+1}^{\al _{k+1}} \cdots \partial _p^{\al _p}
\sigma _j), 
\label{eq:clofX_kofees:ma-conservationlaws}
\ee 
where $1\le k\le p$, $\al =(\al _1,\cdots,\al _p)$, 
and an empty sum is understood to be zero. 
Therefore,  $T=\sigma ^T\rho $
 is a conserved density 
of the system
(\ref{eq:systemofEEs:ma-conservationlaws}),
and $X=(X_1,\cdots,X_p)^T\in {\cal B}^p$ is the conserved flux vector of 
the system
(\ref{eq:systemofEEs:ma-conservationlaws}), corresponding to 
$T$.
\end{theorem}

{\it Proof:}
A simple application of Theorem \ref{thm:gclstructure:ma-conservationlaws} 
to the case of $\Delta =u_t-K(u)$ with a set of independent variables $(t,x^1,\cdots,x^p)$ presents the result in the above thorem. 
  
Alternatively, we can directly prove the theorem. First, we can have  
\bea  &&
(\sigma ^T\rho )_t=\sigma ^T_t\rho + \sigma ^T\rho_t 
=\rho^T\sigma _t+ \sigma ^T\rho_t
= 
  \rho^T K'\sigma - \sigma ^T(K^{'})^*\rho  
\nonumber \\
&=&
\sum_{i,j=1}^q ( \rho_iV_j(K_i)\sigma_j- \sigma _j V_j^* (K_i)
\rho_i ) =
\sum_{i,j=1}^q \sum_{\al \ge 0} ( \rho_i\frac {\part K_i}{\part u^j_\al }D^\al \sigma _j - 
\sigma _j (-D)^\al \rho_i \frac {\part K_i}{\part u^j_\al } ).
\nonumber \eea
Then,
together with Lemma \ref{le:basiceq:ma-conservationlaws}
for all $ 1\le i,j\le q $ and $\al \ge 0$, we can see that 
\eqref{eq:clofees:ma-conservationlaws} holds for all solutons of 
(\ref{eq:systemofEEs:ma-conservationlaws})
and 
the conserved flux vector $X$ is given by \eqref{eq:clofX_kofees:ma-conservationlaws}.  
The proof is finished.
$\vrule width 1mm height 3mm depth 0mm$

The theorem gives us a direct formulation of conservation laws 
for systems of evolution equations, and 
 all expressions are explicitly given for 
the conserved density and conserved fluxes.
The involved conserved density is just a product of a symmetry and an adjoint symmetry,
but the conserved fluxes are dependent on 
the pair of a symmetry and an adjoint symmetry
and the system itself.

\subsection{Recursion structures}

For systems of evolution equations, we can easily prove the following theorem, which states sufficient 
conditions for being recursion operators, adjoint recursion operators, Noether
operators or inverse Noether operators.

\begin{theorem} 
The operators $\Phi (x,t,u) $, $ \Psi(x,t,u)$, $\bar \Phi (x,t,u)$ or $\bar \Psi (x,t,u)$
are a recursion operator, an adjoint recursion operator, a Noether
operator or an inverse Noether
operator of the system 
(\ref{eq:systemofEEs:ma-conservationlaws}),  
if they satisfy 
\bea && 
\frac{\part \Phi }{\part t}+ \Phi '[K]+[\Phi ,K']=0,\\
&& \frac{\part \Psi }{\part t}+\Psi '[K]+[(K^{'})^*,\Psi ]=0 ,\\
&&\frac {\part {\bar \Phi}}{\part t}+{\bar \Phi}'[K]-{\bar \Phi }(K^{'})^*-K'{\bar \Phi }
=0,
\\
&&
 \frac {\part {\bar \Psi}}{\part t}+ {\bar \Psi} '[K]
+ {\bar \Psi} K'+ (K^{'})^*
{\bar \Psi }=0,
 \eea
respectively, where $K'$ and $(K^{'})^*$ denote the Gateaux operator of $K$ and 
its adjoint operator, and the Gateaux
operator of an operator $A$ is similarly defined by 
\be A'[K]=A'(u)[K]=\frac {\part }{\part \varepsilon }\Bigl.\Bigr|_{\varepsilon =0} 
A(u+\varepsilon K) .
  \ee 
\end{theorem}

When the operators $\Phi,\,\Psi ,\, \bar \Phi ,\,\bar \Psi $ 
don't explicitly depend on $t$, the above theorem presents
the results established in \cite{FuchssteinerF-PD1981}.    
It is also easy to see that we have two relations 
\be \Phi ^* =\Psi,\ \bar \Phi ^{-1}=\bar \Psi , \ee 
which means that the adjoint operator of 
a recursion operator (or an adjoint recursion operator) of 
(\ref{eq:systemofEEs:ma-conservationlaws})  
is an adjoint recursion operator (or a recursion operator) of the same system,
and that the inverse operator of a Noether operator (or an inverse Noether operator) of 
(\ref{eq:systemofEEs:ma-conservationlaws})  
is an inverse Noether operator (or a Noether operator) 
of the same system.
It is also easy to prove that
the inverse operator $\Phi^{-1}$ (or $\Psi ^{-1}$) of a recursion operator $\Phi$ (or 
an adjoint recursion operator $\Psi $) of 
a system of evolution equations
is still a recursion operator (or an adjoint recursion operator)
of the same system, if it exists.

\begin{definition}
If a conserved density
$T\in {\cal B}$ 
of a system of evolution equations 
(\ref{eq:systemofEEs:ma-conservationlaws}) is equivalent to zero function,
then $T$ is called trivial, and otherwise $T$ is called nontrivial. 
\end{definition}

Any spatial divergence function 
$\sum_{i=1}^p\part _if_i$ 
must be a trivial conserved density of any system of evolution equations. 
According to the definition \eqref{eq:innerproducts:ma-conservationlaws} of adjoint operators, 
we can have 
\be (\Phi^i\sigma)^T(\Psi^j\rho)\sim \sigma ^T(\Psi^{i+j}\rho)\sim (\Phi^{i+j}\sigma)^T\rho,\ i,j\ge 0,
\label{eq:equivalentconditionforPhi^isigmaPsi^jrho:ma-conservationlaws} \ee
where $\Psi =\Phi^*$.
The study of conservation laws aims at presenting nontrivial conserved densities.
Thus, we only need to consider 
$\sigma ^T(\Psi ^i\rho)$, $(\Phi ^i\sigma )^T\rho$, $i\ge 0$,  
among $(\Phi ^i\sigma )^T(\Psi ^j\rho )$, $i,j\ge 0$, in order to generate 
nontrivial and nonequivalent conserved densities, while using recursion operators.
 
For the Korteweg-de Vries (KdV) equation 
\be u_t=K(u)=\frac 14 u_{xxx}+\frac 32 uu_x,\ x,t\in \R, \label{eq:KdV:ma-conservationlaws} \ee 
it is easy to get  
\be K'=\frac 32 u_x+\frac 32u\part _x +\frac 14\part ^3_x,\ 
(K^{'})^*=-\frac 32u\part _x-\frac 14\part _x^3, \ \part_x=\frac{\part }{\part x},\ee 
and then a recursion operator, an adjoint recursion operator, 
a Noether operator, and an inverse Noether 
operator of the KdV equation (\ref{eq:KdV:ma-conservationlaws}):
\be \Phi = \frac 12 u_x\part _x^{-1}+u +\frac 14\part _x^2, \ \Psi
=-\frac 12 \part _x^{-1}u_x+u+\frac 14 \part _x^2,\ 
\bar \Phi =\part _x,\ \bar \Psi=\part _x^{-1}. 
\label{eq:defofPhiPsi:ma-conservationlaws}
\ee 
Therefore, the KdV equation (\ref{eq:KdV:ma-conservationlaws}) possesses infinitely many 
symmetries and adjoint symmetries, $\Phi ^i\sigma $ and $ \Psi^i\rho$, $i\ge 0$, once 
we have a pair of a symmetry $\sigma$ and an adjoint symmetry $\rho$. We will see later that this really happens. 

\section{Illustrative examples}
\setcounter{equation}{0}

Now we go on to illustrate by examples rich structures of the conserved densities
resulting from symmetries and adjoint symmetries. 

\subsection{The heat equation} 

Let us consider the heat equation
\be u_t=K(u)=u_{xx} ,\ x,t\in \R,\label{eq:he:ma-conservationlaws}\ee 
which lacks a Lagrangian formulation.
Its linearized equation and adjoint linearized equation
read as 
\bea 
&& \sigma_t=K'\sigma=\sigma_{xx}, \\ 
&& \rho_t=-(K^{'})^*\rho=-\rho_{xx},
\eea
respectively. 

Among the first-order differential operators,
it is easy to obtain two recursion operators of (\ref{eq:he:ma-conservationlaws}):
\be \Phi_1=\part _x,  \ \Phi_2=2t\part _x+x, \ee 
and thus two adjoint recursion operators of (\ref{eq:he:ma-conservationlaws}):
\be \Psi_1 =\Phi_1^* = -\part _x,\   \Psi_2 =\Phi_2^*= -2t\part _x+x.\ee
Obviously, 
any solution $f=f(x,t)$ to $f_t=f_{xx}$ and 
the function $K_0=u$ are symmetries of 
the heat equation (\ref{eq:he:ma-conservationlaws}), and 
any solution $g=g(x,t)$ to $g_t=-g_{xx}$ and 
the function $S_0=u(x,-t)$ are adjoint symmetries of 
the heat equation (\ref{eq:he:ma-conservationlaws}). Here the adjoint symmetry $S_0$ has a local
dependence on $u$ with respect to $x$, 
similar to the symmetry $K_0$. 
Only for linear equations, adjoint symmetries 
can have such a local dependence on $u$.

Now, by the principle in Theorem \ref{thm:clstructureforees:ma-conservationlaws}, we have 
infinitely many conserved densities
\be 
g(x,t)u(x,t),
 \ f(x,t)u(x,-t), \ 
 u(x,-t) (\Phi _2^i\Phi _1^j u )(x,t),\ i,j\ge 0,
\label{eq:conserveddensitiesofheatequation:ma-conservationlaws}\ee 
besides a class of trivial conserved densities given by $fg$.
Noting that we have (\ref{eq:equivalentconditionforPhi^isigmaPsi^jrho:ma-conservationlaws})
and 
\be [\Phi_1,\Phi _2]=\Phi_1\Phi_2-\Phi_2\Phi_1=1,\ee 
we did not list the other equivalent or linear conbination type
conserved densities
such as 
\[\ba {l}
(\Psi_1^ig)(\Phi _2^j\Phi _1^k u)\sim (\Psi _1^k\Psi_2^j\Psi_1^i g)u={\bar g}u,\vspace{2mm}
\\  
u(x,-t)(\Phi_1\Phi_2\Phi_1 u)(x,t) =u(x,-t)(\Phi_2\Phi_1^2 u)(x,t)+u(x,-t)(\Phi_1u)(x,t),  
\ea \]
where ${\bar g}$ is a new solution to $g_t=-g_{xx}$.
The first class of conserved densities $gu$ in
 (\ref{eq:conserveddensitiesofheatequation:ma-conservationlaws}) presents
the ones generated in \cite{AncoB-PRL1997},
and the third class of conserved densities contains 
the following interesting conserved densities: 
\bea && 
u(x,-t)(\Phi_2u)(x,t)= u(x,-t) (xu(x,t)+2tu_x(x,t)),\\
&&  u(x,-t)(\Phi_2^2u)(x,t)=u(x,-t)[(x^2+2t)u(x,t)+4txu_x(x,t)+4t^2u_{xx}(x,t)].\qquad
\eea 

\subsection{Burgers' equation}

Let us now consider Burgers' equation
\be u_t=K(u)=2uu_x+u_{xx},\ x,t\in \R, \label{eq:Burgerse:ma-conservationlaws}\ee 
which lacks a Lagrangian formulation as well.
Its linearized equation and adjoint linearized equation
read 
\bea 
&& {\tilde \sigma}_t=K'{\tilde \sigma}= 2u{\tilde \sigma}_x+2u_x{\tilde \sigma} +
{\tilde \sigma}_{xx}, \\ 
&& {\tilde \rho}_t=-(K^{'})^*{\tilde \rho}=2 u{\tilde \rho}_x-{\tilde \rho}_{xx},
\eea
respectively. 

Since the Cole-Hope transformation 
\be v=B(u)=\textrm{e}^{\part _x^{-1}u} \ \textrm{or}\ 
u=(\ln v)_x
\ee
linearizes Burgers' equation (\ref{eq:Burgerse:ma-conservationlaws})
to the heat equation (\ref{eq:he:ma-conservationlaws}) with a dependent variable $v$,
we can move all above results for the heat equation to Burgers' equation.
Note that the Gateaux operator of $B$ and its inverse operator read  
\be B'=\textrm{e}^{\part _x^{-1}u} \part _x^{-1},\ (B')^{-1}=\part _x
\textrm{e}^{-\part _x^{-1}u}.\ee 
The Cole-Hope transformation gives us two recursion operators
for Burgers' equation (\ref{eq:Burgerse:ma-conservationlaws}):
\be \left \{\ba {l}
{\tilde \Phi}_1 =(B')^{-1}\Phi_1B'=
\part _x\textrm{e}^{-\part _x^{-1}u}\part _x
\textrm{e}^{\part _x^{-1}u}\part _x^{-1}=u_x\part _x^{-1}+u +\part _x,\vspace{2mm}\\
{\tilde  \Phi}_2=(B')^{-1}\Phi_2B'=
 \part _x\textrm{e}^{-\part _x^{-1}u}(2t\part _x+x)
\textrm{e}^{\part _x^{-1}u}\part _x^{-1}
=2t {\tilde \Phi}_1 + x+\part _x^{-1},
\ea \right.
\label{eq:mostgeneralsymmetriesofBurgerse}
\ee
and two relations on symmetries and adjoint symmetries between two equations
\be {\tilde \sigma }=(B')^{-1}\sigma ,\ {\tilde \rho}=(B')^\dagger \rho. \ee 
Now, the conserved densities by 
the principle in Theorem \ref{thm:clstructureforees:ma-conservationlaws} can be computed 
as follows:
\be {\tilde \sigma }{\tilde \rho}= ((B')^{-1}\sigma )((B')^\dagger \rho)
=(\part _x \textrm{e}^{-\part _x^{-1}u}\sigma )
(- \part _x^{-1}\textrm{e}^{\part _x^{-1}u}\rho) \sim \sigma \rho.
\ee 
Therefore, for example, if we choose 
$\sigma =g$ and $ \rho = v$, where $g$ solves $g_t=-g_{xx}$, then 
we can immediately obtain a class of conserved densities for Burgers' equation 
(\ref{eq:Burgerse:ma-conservationlaws}):
\be h_0= gv=g \textrm{e}^{\part _x^{-1}u}, \ee 
which was generated in \cite{AncoB-PRL1997}.
In fact, this class of conserved densities corresponds to the following conservation laws
\be h_{0t}=\part _t(g\textrm{e}^{\part _x^{-1}u} )=\part _x(gu
\textrm{e}^{\part _x^{-1}u}-g_x\textrm{e}^{\part _x^{-1}u}). \ee 

Let us set two basic symmetries of Burgers' equation (\ref{eq:Burgerse:ma-conservationlaws}):
\bea && {\tilde K}_0=(B')^{-1}v_x=(B')^{-1}(u\textrm{e}^{\part _x^{-1}u})=
\part _x(\textrm{e}^{-\part _x^{-1}u}u\textrm{e}^{\part _x^{-1}u})=
u_x,\\ &&
{\tilde f}_0=(B')^{-1}f=\part _x(\textrm{e}^{-\part _x^{-1}u}f)=
(f_x-fu)\textrm{e}^{-\part _x^{-1}u},
 \eea 
where $f$ solves $f_t=f_{xx}$.
Since there exist two inverse recursion operators 
 \cite{Li-SCA1990}:
\be 
{\tilde \Phi}_1 ^{-1}= \part _x \textrm{e}^{-\part _x^{-1}u}\part _x^{-1}
\textrm{e}^{\part _x^{-1}u}\part _x^{-1}
 ,\  
 {\tilde \Phi}_2^{-1}= 
\part _x\textrm{e}^{-\part _x^{-1}u}(2t\part _x+x)^{-1}
\textrm{e}^{\part _x^{-1}u}\part _x^{-1}
, \ee 
where the inverse in the middle of the second formula can be worked out:
\[(2t\part _x+x)^{-1}=
\frac 1{2t}\textrm{e}^{-\frac {x^2}{4t}}\part _x^{-1}\textrm{e}^{\frac {x^2}{4t}},\]
we can have infinitely many symmetries for Burgers' equation (\ref{eq:Burgerse:ma-conservationlaws}): 
%\cite{Li-SCA1990}
\be {\tilde K}_{ij}={\tilde \Phi}_2 ^i{\tilde \Phi}_1^j{\tilde K}_0,\ 
{\tilde f}_{ij}={\tilde \Phi}_2 ^i{\tilde \Phi}_1^j{\tilde f}_0,
%(2tu_x+1)
\ i,j\in \Z , \label{eq:mostgeneralsymmetriesofBurgerse}\ee
where we can not add the other symmetries such as 
${\tilde \Phi}_1{\tilde \Phi}_2{\tilde \Phi}_1{\tilde K}_0$
to the algebra spanned by 
all symmetries in (\ref{eq:mostgeneralsymmetriesofBurgerse}) because 
\[ [{\tilde \Phi}_1,{\tilde \Phi}_2]={\tilde \Phi}_1{\tilde \Phi}_2 -
{\tilde \Phi}_2{\tilde \Phi}_1=1. \]
The symmetries defined by (\ref{eq:mostgeneralsymmetriesofBurgerse}) 
contain all symmetries generated in \cite{Tian-SCA1988},
and in particular, we have a time-dependent symmetry 
\be {\tilde K}_{1,-1}={\tilde \Phi}_2{\tilde \Phi}_1^{-1}u_x=
(2t {\tilde \Phi }_1+ x+\part _x^{-1}){\tilde \Phi}_1^{-1}u_x
%=2t u_x+(x+\part _x^{-1})(\part _x \textrm{e}^{-\part _x^{-1}u}\part _x^{-1}
%\textrm{e}^{\part _x^{-1}u}\part _x^{-1}u)
=2tu_x+1. \ee 
Moreover, a direct computation can show that 
all local adjoint symmetries of (\ref{eq:Burgerse:ma-conservationlaws}), 
depending on $x,t,u$, and derivatives of $u$ with 
respect to $x$ to some finite order, must be a constant function. 
By noting that 
\[x+\part _x^{-1}=\part _xx\part _x^{-1},\ u_x\part _x^{-1}+u=\part _xu\part _x^{-1}, \]
%Theorem \ref{thm:clstructureforees:ma-conservationlaws} tells us 
all conserved densities resulted from the products of the above symmetries and a local 
adjoint symmetry ${\tilde \rho}_0=1$ must be trivial.
For example, the following class of conserved densities 
\be {\tilde \rho}_0{\tilde f}_0=(f_x-uf)\textrm{e}^{-\part _x^{-1}u}=\part _x(f  \textrm{e}^{-\part _x^{-1}u}) \ee
is trivial. This result also provides an evidence   
why Burgers' equation (\ref{eq:Burgerse:ma-conservationlaws}) has
only one nontrivial conserved density of differential polynomial type
\cite{TuQ-SCA1981}.   

Nevertheless, based on two basic adjoint symmetries of the heat equation, we can obtain
two nonlocal basic adjoint symmetries of Burgers' equation (\ref{eq:Burgerse:ma-conservationlaws}), defined by
\be {\tilde g}_0=(B')^\dagger g = 
-\part _x^{-1}(g\textrm{e}^{\part _x^{-1}u}),
\ {\tilde S}_0=(B')^\dagger S_0=-\part _x^{-1}
(S_0\textrm{e}^{\part _x^{-1}u}),
 \ee 
where $g$ solves $g_t=-g_{xx}$ and $S_0(x,t)=u(x,-t)$.
Now, by the principle in Theorem \ref{thm:clstructureforees:ma-conservationlaws}, we have 
infinitely many conserved densities
\be {\tilde g}_0{\tilde K}_{ij}
%={\tilde g}_0({\tilde \Phi}_2^i{\tilde \Phi}_1^j{\tilde K}_0)
, \ {\tilde g}_0{\tilde f}_{ij}
%={\tilde g}_0({\tilde \Phi}_2^i{\tilde \Phi}_1^j{\tilde f}_0)
, \ {\tilde S}_0{\tilde K}_{ij}
%={\tilde S}_0({\tilde \Phi}_2^i{\tilde \Phi}_1^j{\tilde K}_0)
, \ {\tilde S}_0{\tilde f}_{ij}
%={\tilde S}_0({\tilde \Phi}_2^i{\tilde \Phi}_1^j{\tilde f}_0)
,  \ i,j\in \Z .
 \label{eq:conserveddensitiesofBurgersequation:ma-conservationlaws}\ee 
Several simple classes of conserved densities can be computed as follows:
\bea && {\tilde g}_0{\tilde f}_0=-\part _x^{-1}
(g\textrm{e}^{\part _x^{-1}u})\part _x(f\textrm{e}^{\part _x^{-1}u}) 
\sim fg :=h_1, \nonumber\\ 
&&{\tilde g}_0{\tilde K}_{0}=-
\part _x^{-1}(g\textrm{e}^{\part _x^{-1}u})u_x\sim
gu\textrm{e}^{\part _x^{-1}u}:=h_2,\nonumber\\ &&
{\tilde g}_0{\tilde K}_{1,-1}=-
\part _x^{-1}(g\textrm{e}^{\part _x^{-1}u})(2tu_x+1) \sim
g(2tu+x) \textrm{e}^{\part _x^{-1}u}:=h_3,\nonumber\\ &&
{\tilde g}_0{\tilde f}_{01}=
-\part _x^{-1}(g\textrm{e}^{\part _x^{-1}u})[\part _x^2(f \textrm{e}^{-\part _x^{-1}u})+
\part _x(fu\textrm{e}^{-\part _x^{-1}u})]\nonumber \\ &&
\qquad \qquad 
 \sim (g \textrm{e}^{\part _x^{-1}u})\part _x(f \textrm{e}^{-\part _x^{-1}u})+fgu=
f_xg :=h_4,\nonumber\\
&& {\tilde g}_0 {\tilde K}_{01}=
-\part _x^{-1}(g\textrm{e}^{\part _x^{-1}u})\part _x(u_x+u^2)\sim
g(u_x+u^2)\textrm{e}^{\part _x^{-1}u}:=h_5.\nonumber
\eea 
They correspond to the following conservation laws
\bea 
&& h_{1t}=\part _t(fg)=\part _x(f_xg-fg_x),\nonumber\\ 
&& 
h_{2t}=\part _t( gu\textrm{e}^{\part _x^{-1}u})
=\part _x[( gu^2+gu_x-g_xu) \textrm{e}^{\part _x^{-1}u}],\nonumber\\
&& 
h_{3t}=\part _t[(2tu+x) g\textrm{e}^{\part _x^{-1}u}]=
\part _x \{[g_x(2tu+x)-g(2tu_x+1)
-g(2tu+x)u]\textrm{e}^{\part _x^{-1}u}\},\qquad \nonumber\\ &&
h_{4t}=\part _t(f_xg )=\part _x(f_{xx}g-f_xg_x) ,\nonumber\\ &&
h_{5t}= \part _t [g(u_x+u^2)\textrm{e}^{\part _x^{-1}u}]=\part _x\{[-g_x(u_x+u^2)+g(
u_{xx}+3uu_x+u^3)]\textrm{e}^{\part _x^{-1}u}\},\qquad \nonumber
\eea 
the first and the fourth of which are trivial conservation laws of the second kind
(see \cite{Olver-book1986} for the definition).
Of course, there are the other two classes of conserved densities, generated from 
the second basic adjoint symmetry ${\tilde S}_0$.  

\subsection{The Korteweg-de Vries equation}

Let us finally consider the Korteweg-de Vries (KdV) equation (\ref{eq:KdV:ma-conservationlaws}), i.e.,
\[ u_t=K(u)=\frac 32uu_x+\frac 14 u_{xxx},\ x,t\in \R. \]
This standard form of the KdV equation also lacks a Lagrangian formulation.
Its linearized equation and adjoint linearized equation
read 
\bea 
&& \sigma_t=K'\sigma= \frac 32u\sigma_x+\frac 32u_x\sigma +\frac 14\sigma_{xxx}, \\ 
&& \rho_t=-(K^{'})^*\rho= \frac 32 u\rho_x + \frac 14\rho_{xxx},
\eea
respectively.
It is known that we have infinitely many symmetries 
\be K_i=\Phi ^iu_x=\part _x\frac{\delta {\tilde H}_i}{\delta u},\ 
\tau _i=\Phi ^i (\frac 32 tu_x+1),
\  {\tilde H}_i= 
\int H_idx,\ H_i\in {\cal A},\ i\ge 0, \ee
with $\Phi $ being given by (\ref{eq:defofPhiPsi:ma-conservationlaws}).
%\[ \Phi = \frac 12 u_x\part _x^{-1}+ u +\frac 14 \part _x^2.\]
They form a Virasoro algebra (see, e.g., \cite{Ma-JPA1990}):
\be 
[K_i,K_i]=0, \ [K_i,\tau_j]=(i+\frac12  )K_{i+j-1}, \ [\tau _i,\tau _j]=(i-j)\tau _{i+j-1},
\ i,j\ge 0,\label{eq:VirasoroaofKdV:ma-conservationlaws}
\ee
where $K_{-1}=\tau _{-1}=0$,
% and $\tau _i$, $ i\ge 0$, are defined by  \[\tau_i=\Phi ^i \part _xS_0=\Phi ^i(\frac 32 tu_x+1),\ i\ge 0,\]
and the commutator is defined by 
\[ [Y,Z]=\frac {\part }{\part \varepsilon } \Bigl.\Bigr|_{\varepsilon =0}(Y(u+\varepsilon Z)
-Z(u+\varepsilon Y)),\ Y,Z\in {\cal B}^q.\]
In particular, we have  
\[ \tau _1=\Phi \tau _0=t(\frac 38u_{xxx}+\frac 64 uu_x)+(\frac 12xu_x+u), \]
but the other time-dependent symmetries $\tau _i,\ i\ge 2$, are all nonlocal.

Since we have an inverse Noether operator $\part _x^{-1}$,  
we can immediately obtain two local adjoint symmetries of the KdV equation (\ref{eq:KdV:ma-conservationlaws}):
\be S_0=\part _x^{-1}K_0=u,\ T_0=\part _x^{-1}\tau _0=\frac 32 tu+x.  \ee
First, note that we have (\ref{eq:equivalentconditionforPhi^isigmaPsi^jrho:ma-conservationlaws}).
%\[(\Phi ^if) (\Psi ^jg) \sim f( \Phi ^{i+j}g),\ f,g\in {\cal B}_c,\ 
%\Psi =\Phi ^* ,\ i,j\ge 0 . \]
Hence, by the principle in Theorem \ref{thm:clstructureforees:ma-conservationlaws}, 
all conserved densities of the type $(\Psi ^i\rho )(\Phi ^j\sigma )$ with
$\Psi=\Phi^*$ and with
$\sigma$ and $\rho$ being the above symmetries and adjoint symmetries
give us the conserved densities
\be S_0\Phi ^iK_0,\ T_0\Phi ^i K_0,\ S_0\Phi ^i \tau _0,\ T_0\Phi ^i\tau_0,
\  i\ge 0. \label{eq:conseveddensitiesofKdV:ma-conservationlaws}\ee 
Second, note that there exist functions $P_i\in {\cal A},\ i\ge 0,$ such that 
\[u\part _x \frac{\delta {\tilde H}_i}{\delta u}=\part _xP_i,\ i\ge 0,\]
and we have a relation among $H_i$:
\[ \frac{\delta {\tilde H}_i}{\delta u}=(i+\frac 12 )H_{i-1},\ H_{-1}=2u,\ i\ge 0,\]
which can be found through the Virasoro algebra (\ref{eq:VirasoroaofKdV:ma-conservationlaws}).
Therefore, we can compute that 
\bea && S_0\Phi ^iK_0= \part _xP_i,\ 
T_0\Phi ^iK_0= -\frac{\delta {\tilde H}_i}{\delta u}
+\part _x (\frac 32 tP_i+x\frac{\delta {\tilde H}_i}{\delta u})\nonumber \\ &&
\quad 
 =-(i+\frac12  ) H_{i-1}+\part _x(\frac 32 tP_i +x\frac{\delta {\tilde H}_i}{\delta u})
, \ i\ge 0,
 \nonumber \\ 
&& S_0\Phi ^i\tau _0 =S_0\Phi ^i\part _xT_0= S_0\part _x\Psi ^iT_0\sim
(\part _xS_0)\Psi ^iT_0=K_0\Psi ^iT_0 \sim T_0\Phi ^iK_0,\ i\ge 0,\nonumber \\ &&
T_0\tau _0=\part _x(\frac 98t^2u^2 +\frac 32 txu+\frac 12 x^2),
 \nonumber \\ &&
T_0\tau _1= \part _x[\frac 94 t^2 P_1+\frac 14 xu^2 -\frac 14 u_x+x(\frac 34 u^2
+\frac 14 u_{xx})+\frac 12 x^2 u].
\nonumber 
\eea
Now it follows that all nontrivial conserved densities defined by 
(\ref{eq:conseveddensitiesofKdV:ma-conservationlaws}) are 
infinitely many nontrivial conserved densities $\{H_i\}_{i=-1}^\infty$, and 
infinitely many nonlocal conserved densities $T_0\Phi ^i\tau_0$, $i\ge 2$.

Lastly, we would like to show 
that  
the Lax pair of the KdV equation can be used to generate 
nonlocal conserved densities.
It is known that 
there exist
 many nonlocal symmetries 
generated from eigenfunctions and adjoint eigenfunctions of the Lax pair \cite{Ma-JPSJ1995}.
Let us just state the main results. 
The KdV equation (\ref{eq:KdV:ma-conservationlaws}) has a Lax pair
\be 
U(u,\la )=\left(\begin{array} {cc}0&1\vspace{2mm} \\ \lambda -u & 0 \end{array} \right)
,\ V(u,\la )= \left(\begin{array} {cc}
-\frac 14 u_x&\lambda +\frac12 u\vspace{2mm} \\ \lambda^2 -\frac 12
u\lambda -\frac 14 u_{xx}-\frac 12 u^2 & \frac 14 u_x \end{array} \right) 
. \ee 
Using this Lax pair, we introduce $N$ replicas of the Lax systems
\[ \phi^{(s)}_x=U(u,\la _s)\phi^{(s)}, \ \phi^{(s)}_t=V(u,\la _s)\phi^{(s)},\
\ \phi^{(s)}=(\phi_{1s},\phi_{2s})^T,\ 1\le s\le N, \] 
and $N$ replicas of the adjoint Lax systems
\[ \psi^{(s)}_x=-U^T(u,\la _s)\psi^{(s)}, \ \psi^{(s)}_t=-V^T(u,\la _s)\psi^{(s)}
,\ \psi^{(s)}=(\psi_{1s},\psi_{2s})^T,
\ 1\le s\le N, \]
where $\la _1,\cdots ,\la _N$ are arbitrary constants.
Then, we have an adjoint symmetry and a symmetry 
\begin{equation}
\rho_0 = P_1^TQ_2, \ \sigma_0=  (P_1^TQ_2)_x= P_1^TQ_1 -P_2^TQ_2, \end{equation}
where $P_i$ and $Q_i$ are $N$ dimensional vector functions, 
\[
 P _i=(\phi _{i1},\phi _{i2},\cdots,\phi
_{iN})^T,\ Q_i=(\psi_{i1},\psi_{i2},\cdots,\psi_{iN})^T ,\ i=1,2 .\]
Therefore, we can have infinitely many conserved densities 
in terms of eigenfunctions and adjoint eigenfunctions:
\be 
%\rho_0K_i= -\sigma_0 \frac {\delta H_i}{\delta u}+\part _x(\rho_0 \frac {\delta H_i}{\delta u}),
\rho_0 \Phi ^i K_0,\ 
\rho_0 \Phi ^i \tau _0, 
\  i\ge 0,
\ee 
the first two conserved densities of which are the following 
\bea && \rho_0K_0=u_x\rho_0 \sim u\sigma _0=u\sum_{s=1}^N(\phi_{1s}\psi_{1s}-\phi_{1s}\psi_{2s})
, 
\nonumber
\\
&&
\rho_0\tau_0\sim \frac 32 tu\sigma _{0}+\rho_0=\frac 32 tu
\sum_{s=1}^N(\phi_{1s}\psi_{1s}-\phi_{1s}\psi_{2s})
+ \sum_{s=1}^N\phi_{1s}\psi _{2s}. \nonumber \eea 

\section{Conclusion and remarks}
\setcounter{equation}{0}

We have established a direct formulation of conservation laws 
for systems of differential equations, regardless of the existence of a Lagrangian, and made an application to systems of evolutions equations, together with three examples of scalar evolution equations. 
The presented examples include computations of 
conserved densities for the heat equation, Burgers' equation,
and the Korteweg-de Vries (KdV) equation.

We remark that 
pairs of symmetries and adjoint symmetries 
have also been used to formulate conservation laws for systems of discrete evolution equations \cite{Ma-Symmetry2015}, and
Theorem \ref{thm:gclstructure:ma-conservationlaws}
could be obtained from 
an application of 
Noether's theorem to
an enlarged Euler-Lagrange system
\be
\frac {\delta {\cal L}}{\delta v}=0, \ \frac {\delta {\cal L}}{\delta u}=0, 
\ee 
where ${\cal L}=\int v^T\Delta (u) \,dx $ with $v=(v^1,\cdots,v^l)^T$.  
%The conservation law in Theorem \ref{thm:gclstructure:ma-conservationlaws}
%is just the conservation law determined by Noether's theorem from a symmetry %$\sigma $, 
%while an adjoint symmetry $v=\rho$ exists.   
The first subsystem is precisely the original system $\Delta =0$, and   
the second subsystem $\frac {\delta {\cal L}}{\delta u}=(\Delta ')^*v=0$ is satisfied 
if we take $v$ as an adjoint symmetry $\rho(u)$ of the system $\Delta =0$. 
When $v=\rho=u$ (or $v=\rho(x,u)$), the system under consideration is called 
strictly (or nonlinearly) self-adjoint,  and the resulting conservation law presents 
the one in 
\cite{Ibragimov-JMAA2007,Ibragimov-JPA2011}. 
In the case of self-adjoint systems of differential equations,
Theorem \ref{thm:gclstructure:ma-conservationlaws}
generates trivial conservation laws of the second kind,
since symmetries are adjoint symmetries, too \cite{AncoB-JMP1996}. 
The idea of constructing conserved densities 
by symmetries and adjoint symmetries 
is also similar to that of binary symmetry constraints, which results in a binary nonlinearization theory \cite{MaS-PLA1994}.  

There exists a differential geometric formulation for attempting adjoint symmetries of the second-order ordinary differential equations 
\cite{MorandoP-JPA1995}, and  
%But we have not used any geometrical notation even for the continuous case.  
adjoint symmetries are also used to show separability of 
Hamiltonian systems of ordinary differential equations
\cite{SarletR-JMP2000}. 
The existence of symmetry algebras are due to 
Lie algebraic structures associated with
 Lax operators corresponding to symmetries
(see \cite{Ma-JPA1992,Ma-JMP1992,Ma-book1993} for systems of continuous evolution equations and 
\cite{FuchssteinerM-book1999,MaF-JMP1999} for systems of discrete evolution equations).
How about 
Lie algebraic structures for adjoint symmetries?  
What kind of commutators between adjoint symmetries can be introduced?
A good candidate for commutators could
be taken as 
\be \left[\!\right[ \rho_1,\rho_2 \left ]\!\right ]=(\rho'_1)^{*}\rho_2-(\rho'_2) 
^{*}\rho_1, \label{eq:adjointbinaryoperation:ma-conservationlaws}\ee
where $(\rho'_1)^*$ and 
$(\rho'_2)^*$ denote the adjoint operator of their Gateaux operators. 
However, this doesn't keep the space of adjoint symmetries closed.
This can be seen from an example in the case of the KdV equation.
The KdV equation (\ref{eq:KdV:ma-conservationlaws}) has 
 two adjoint symmetries $\rho_1=u$ and $\rho_2=\frac 32 tu+x$, whose expected commutator reads
\be  \left[\!\right [ \rho_1,\rho_2 \left ]\!\right ]= \left[\!\right [u, 
\frac 32 tu+x\left ]\!\right ]= x. \ee 
But the resulting function $x$ is not an adjoint symmetry of the KdV equation (\ref{eq:KdV:ma-conservationlaws}).
The characteristic in \eqref{eq:Qforsigmaandrho:ma-conservationlaws} may be useful in exploring a successful commutator of adjoint symmetries. 

\vspace{2mm} 

{\bf Acknowledgment}
This work was in part supported by 
NSF under the grant DMS-1664561,
 NNSFC under the grants 11371326, 11371086 and 11571079, 
the 111 Project of 
of China (B16002), 
and the distinguished professorships of the Shanghai University of Electric Power and Shanghai Second Polytechnic 
University.

\end{document}